\documentclass{ICSnewsletter}

\usepackage[utf8]{inputenc}
\usepackage{amsfonts}
\usepackage{graphicx}
\usepackage{xcolor}
\usepackage{subcaption}
\usepackage{hyperref}

\usepackage{amsmath}
\usepackage{adjustbox}
\usepackage[normalem]{ulem}
\usepackage{mathtools}
\usepackage{glossaries}

\usepackage{siunitx}

\usepackage{listings}

\usepackage{nicefrac}

\usepackage{bm}

\usepackage{tikz}
\usepackage{tikz-cd}

\usepackage{pgfplots}
\usepgfplotslibrary{colorbrewer}

\newcommand{\vect}[1]{\ensuremath{\bm{\mathbf{#1}}}}
\newcommand{\mat}[1]{\ensuremath{\mathbf{#1}}}

\newcommand{\parder}[2]{\ensuremath{\frac{\partial #1}{\partial #2}}}
\newcommand{\grad}[1]{\ensuremath{\nabla #1}}
\renewcommand{\div}[1]{\ensuremath{\nabla \cdot #1}}
\newcommand{\curl}[1]{\ensuremath{\nabla \times #1}}

\newcommand{\Ltwo}[1]{\ensuremath{L^{2}\left(#1\right)}}

\newcommand{\Hone}[2]{\ensuremath{H^{1}_{#2}\left(#1\right)}}

\newcommand{\Hdiv}[2]{\ensuremath{\mathbf{H}_{#2}\left(\text{div};#1\right)}}
\newcommand{\Hcurl}[2]{\ensuremath{\mathbf{H}_{#2}\left(\textbf{curl};#1\right)}}

\newacronym{cad}{CAD}{Computer Aided Design}
\newacronym{fem}{FEM}{Finite Element Method}
\newacronym{bem}{BEM}{Boundary Element Method}
\newacronym{iga}{IGA}{Isogeometric Analysis}
\newacronym{nurbs}{NURBS}{Non-Uniform Rational B-splines}
\newacronym{pmsm}{PMSM}{Permanent Magnet Synchronous Machine}
\newacronym{rf}{RF}{Radio Frequency}
\newacronym{pde}{PDE}{Partial Differential Equation}

\title{Recent Advances of Isogeometric Analysis in Computational Electromagnetics\\ \small Z. Bontinck, J. Corno, H. De Gersem, S. Kurz, A. Pels, S. Schöps, F. Wolf, C. de Falco, J. Dölz, R. Vázquez and U. Römer}

\begin{document}
\maketitle

\begin{abstract} 
In this communication the advantages and drawbacks of the isogeometric analysis (IGA) are reviewed in the context of electromagnetic simulations. IGA extends the set of polynomial basis functions, commonly employed by the classical Finite Element Method (FEM). While identical to FEM with Nédélec's basis functions in the lowest order case, it is based on B-spline and Non-Uniform Rational B-spline basis functions. The main benefit of this is the exact representation of the geometry in the language of computer aided design (CAD) tools. This simplifies the meshing as the computational mesh is implicitly created by the engineer using the CAD tool. The curl- and div-conforming spline function spaces are recapitulated and the available software is discussed. Finally, several non-academic benchmark examples in two and three dimensions are shown which are used in optimization and uncertainty quantification workflows.
\end{abstract}

\section{Introduction}\label{sec:intro}
In electrical engineering numerical simulations have become an inevitable tool for designing new components, such as electrical machines or antennas,  as well as for understanding their behavior and their interaction with the environment, as e.g. commonly analyzed in electromagnetic compatibility simulations. However, due to exponentially increasing computational resources and increasing accuracy expectation of the simulation engineers, the numerical models become correspondingly more and more complex.

There are many trends leading to this increased complexity and the following list is obviously not complete. For example, multiphysical phenomena are considered to be increasingly important and they lead to many challenges in the modeling and the numerical analysis, see e.g. \cite{Clemens_2012ab}. On the other hand, the geometrical design of components to be simulated has become very complex. It is often directly imported from a \gls{cad} software into the engineering tools that are the main interest of the Compumag community. Geometrical simplifications of the design, e.g.~removing screws, require a lot of manual labor and sometimes even insights into the actual field distribution within the component, which should have been gained by the simulation in the first place. According to \cite{Hughes_2005aa}, approximately \SI{80}{\percent} of the overall analysis time nowadays can be accounted to mesh generation in the automotive, aerospace, and ship building industries. At the same time, the higher accuracy demands require even finer meshes or higher order ansatz functions, which in turn necessitate the creation of meshes with curved elements. Now, if the initial mesh generation for realistic computational models is already troublesome, how can one robustly carry out shape optimization or uncertainty quantification of geometrical manufacturing imperfections?

In this communication, we advertise \gls{iga} as proposed by Tom Hughes in \cite{Hughes_2005aa}, see e.g.~\cite{Cottrell_2009aa}. Its distinct feature is to mitigate the meshing step and immediately work on the \gls{cad} geometry description. The idea of more sophisticated parametric, i.e.~non-polynomial, mappings is not new but gained further interest also by \gls{cad} software vendors due to the recent works on \gls{iga}. Eventually, the mesh shall be automatically created by the engineer using the \gls{cad} software. Currently, this is not entirely true as boundary representations (\emph{BRep}) of volumetric objects are used in most \gls{cad} tools. The trivariate mappings of the interior have still to be created  manually. However, this is an active and ongoing field of research \cite{Akhras_2016aa}.

Eventually, the \gls{fem} is the work horse in low-frequency and many high-frequency electromagnetic simulations \cite{Monk_2003aa,Yin_2014aa}. In the last decades, many variants and improvements of the Finite Element Method have been proposed.
\gls{iga} proposes to generalize the set of polynomial basis functions, employed by \gls{fem}, with the introduction of more general B-spline basis functions and \gls{nurbs}. These functions are well known in the design community and are the basic ingredient of nowadays \gls{cad} software. They  allow for the exact parametrization of common curves and surfaces such as conic sections, and are extremely flexible and intuitive when dealing with more complex shape creation and deformation. By using them for the discretization process of general \glspl{pde}, it is possible to inherit these properties and directly use the \gls{cad} design as the computational domain, without the need of a meshing step.

\begin{figure}
	\centering
	\includegraphics[width=0.4\textwidth]{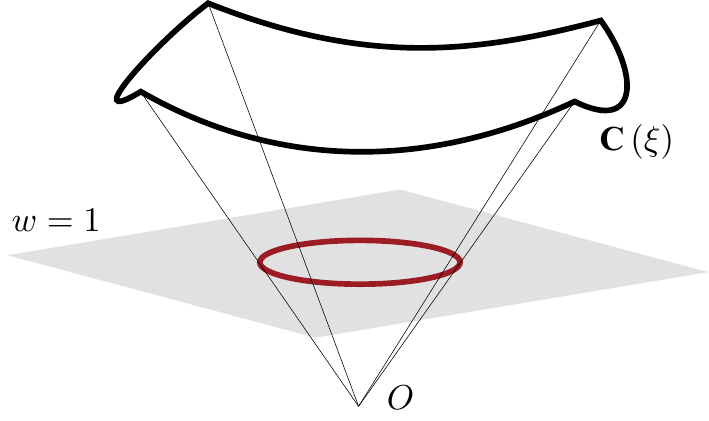}
	\caption{A \gls{nurbs} curve in $\mathbb{R}^2$ is the projection of a B-spline curve in $\mathbb{R}^3$, this allows for the definition of circles and other conic sections.}
	\label{fig:circle}
\end{figure}

It is also worth pointing out that, since \gls{iga} operates in the same Galerkin framework as \gls{fem}, in most instances pre-existent finite element codes can be easily modified in order to work in an isogeometric setting by changing the basis function construction routines only.

The possibility to straightforwardly deal with geometrical changes makes of \gls*{iga} a powerful tool when used in the context of shape optimization problems or shape sensitivity analysis~\cite{Cottrell_2009aa,Wall_2008aa,Qian_2010aa}. Moreover, given the higher global regularity of the basis function space, the isogeometric approach has been shown to present several advantages over \gls{fem} in addition to the better handling of geometries, like a faster convergence with respect to the number of degrees of freedom \cite{Hughes_2008aa} and the possibility to treat high order differential operators \cite{Bartezzaghi_2015aa,Hosseini_2015aa}. 

These properties have made \gls{iga} appealing for a wide variety of applications. For the application to computational electromagnetics, in particular, \gls{iga} shows the ability of consistently discretizing complexes of differential forms \cite{Buffa_2011aa} which is a property of great importance for achieving spectrally correct discretization of the Maxwell PDEs \cite{Buffa_2010aa,Buffa_2013aa}.

In this communication we present a general introduction to the isogeometric framework highlighting some differences, advantages and drawbacks with respect to \gls{fem}. We also introduce the curl- and div-conforming spaces necessary for the correct discretization of Maxwell's equations as given in \cite{Buffa_2010aa}. We then proceed to present several applications of \gls{iga} to different kind of problems both in 2D and in 3D. Some final remarks on an isogeometric boundary element method are also given.
 
\section{Isogeometric Discretization of Maxwell's Equations}\label{sec:discretisation}

The differential form of Maxwell's equations is given by
\begin{subequations}\label{eq:Maxwell}
\begin{align}
\curl{\vect{E}} + \parder{\vect{B}}{t} &= 0\\
\div{\vect{D}} 										     &= \rho\\
\curl{\vect{H}} - \parder{\vect{D}}{t} &= \vect{J}\\
\div{\vect{B}}                       	 &= 0,
\end{align}
\end{subequations}
with $\vect{E}$ and $\vect{D}$ the electric field strength and electric flux density, $\vect{H}$ and $\vect{B}$ the magnetic field strength and magnetic flux density, $\rho$ the electric charge density and $\vect{J}$ the current density. The system is completed by the material relations
\begin{subequations}\label{eq:constit-laws}
\begin{align}
\vect{D} &= \epsilon\vect{E} + \vect{P} \label{eq:constit-law-D}\\
\vect{B} &= \mu\left(\vect{H} + \vect{M}\right) \label{eq:constit-law-B}
\end{align}
\end{subequations}
where the electric permittivity $\epsilon$ and the magnetic permeability $\mu$ are, in general, non-linear tensor valued functions of position expressing the material properties, and the quantities $\vect{P}$ and $\vect{M}$ are the electric polarization and the magnetization respectively.

\begin{figure}
\centering
\includegraphics[width=0.45\textwidth]{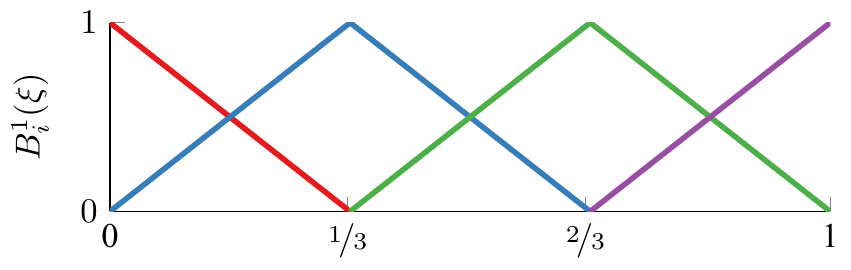}\\
\includegraphics[width=0.45\textwidth]{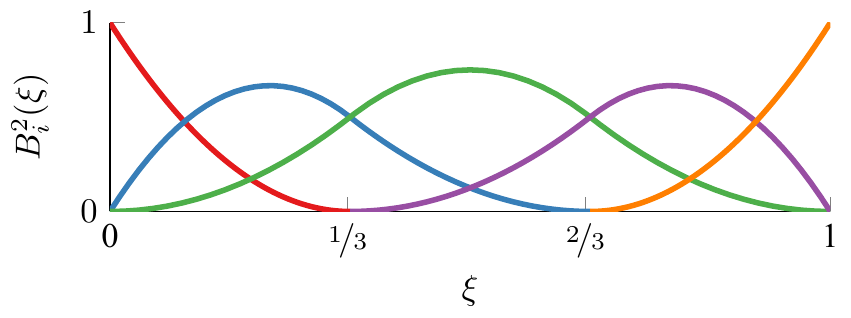}
\vspace{-0.5em}
\caption{B-spline basis functions of degree 1 and 2 on open, uniform knot vectors ($\Xi = \left[0,0,\nicefrac{1}{3},\nicefrac{2}{3},1,1\right]$ on top and $\Xi = \left[0,0,0,\nicefrac{1}{3},\nicefrac{2}{3},1,1,1\right]$ at the bottom).}\label{fig:BSp-basis}
\end{figure}

Depending on the problem at hand, different assumptions and simplifications can be made. We focus here on high-frequency problems in frequency domain and low-frequency problems that can be considered quasi-static or even static. We postpone the detailed discussion of the specific formulations and models to section~\ref{sec:applications}. In general, however, the numerical solution of such problems is nowadays typically performed with the \gls{fem}. The spaces arising from the weak formulation are non-standard spaces of square integrable functions (i.e.~in $L^2$) with weakly defined \textbf{curl} in $L^2$; this space is commonly denoted by $\Hcurl{\cdot}{}$. In order to properly approximate the solution field, the discrete spaces need to mimic several important properties that hold on the continuous level (see sub-section \ref{sec:iga-derham}). For classical polynomial \gls{fem} it is well known that a sequence of discrete spaces with conforming discretization can be obtained using (high-order) Nédélec-type elements, which allocate the degrees of freedom on the edges or the faces of the mesh in order to ensure consistency \cite{Monk_2003aa,Yin_2014aa}.

\subsection{IGA Basis Functions}
In order to represent complex shapes, the use of polynomials or rational segments may often be inadequate or imprecise. On the other hand, B-spline and \gls{nurbs} functions enjoy some major advantages that make them extremely convenient for surface representation and are therefore the most common choice for solid geometry modeling on which nowadays \gls{cad} tools are based. Of the main advantages we care to mention, e.g., that they can exactly represent all conic sections (i.e.~circles, ellipses, etc.), see e.g.~Fig.~\ref{fig:circle}, that they can be generated by many efficient and numerically stable algorithms, and that they can easily handle specified continuity in single points \cite{Blanc_1996aa}.

The main idea behind the isogeometric approach \cite{Hughes_2005aa} is to discretize the problem unknowns with the same set of basis functions that \gls{cad} employs for the construction of geometries. 

Let $p$ be the prescribed degree and
\begin{equation}
\Xi=\begin{bmatrix}
\xi_{1} & \dots & \xi_{n+p+1}
\end{bmatrix}
\end{equation}
be a vector that partitions $[0,1]$ into elements ($\xi_i\in\hat{\Omega} = [0,1]$). Then, the Cox-de Boor's formula~\cite{Piegl_1997aa} defines $n$ one-dimensional B-spline basis functions $\lbrace B_i^p(\xi)\rbrace$ with $\xi\in\hat{\Omega}$ as
\begin{align}\label{eq:Cox-deBoor}
B_i^0\left(\xi\right) &= \begin{cases}
1 & \text{if }\xi_i\leq\xi<\xi_{i+1}\\
0 & \text{otherwise}
\end{cases}\\
B_i^p\left(\xi\right) &= \frac{\xi-\xi_{i}}{\xi_{i+p}-\xi_{i}}B_i^{p-1}\left(\xi\right)+\frac{\xi_{i+p+1}-\xi}{\xi_{i+p+1}-\xi_{i+1}}B_{i+1}^{p-1}\left(\xi\right),\nonumber
\end{align}
with $i=1,\dots,n$. An example of B-spline basis of degree $p=1,2$ are shown in Fig.~\ref{fig:BSp-basis}. One can notice that the support of a B-spline of degree $p$ is always $p+1$ knot spans and, as a consequence, each $p$-th degree function has $p-1$ continuous derivatives across the element boundaries (i.e.~across the knots) if they are not repeated. Repetition of knots can be exploited to prescribe the regularity.
\begin{figure}
\centering
\includegraphics[width=0.45\textwidth]{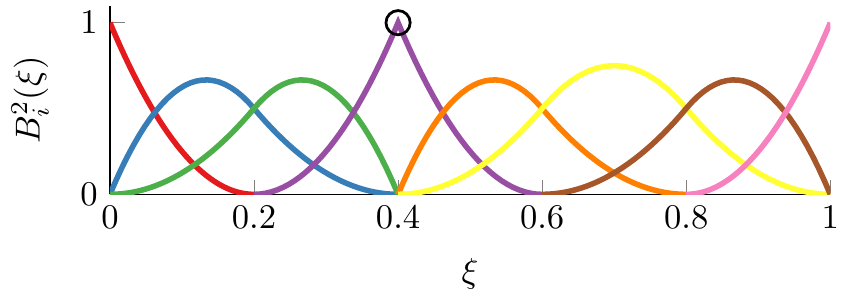}\\
\includegraphics[width=0.45\textwidth]{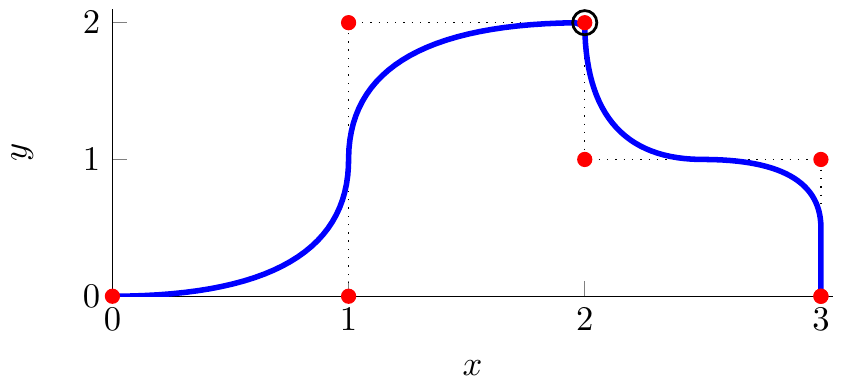}
\caption{Continuity can be controlled in single points (see the black circle) by knot repetition (here: $\Xi = \left[0, 0, 0, .2, .4, .4, .6, .8, 1, 1, 1\right]$).}
\label{fig:bsplines_regularity}
\end{figure}

\gls*{nurbs} of degree $p$ are defined as rational B-splines
\begin{equation}\label{eq:nurbs_basis}
N_i^p(\xi) = \frac{w_i B_i^p(\xi)}{\sum_j w_j B_j^p(\xi)},
\end{equation}
with $w_i$ a weighting parameter associated with the $i$-th basis function. It is clear that B-splines are \gls{nurbs} with all weights equal to one, due to the partition of unity property.

Multi-dimensional B-splines and \gls{nurbs} are constructed using a tensor product approach. For example, let $\Xi_d$ be the knot vectors, $p_d$ the degrees and $n_d$ the number of basis functions (with $d=1,2,3$), a trivariate B-spline is given by
\begin{equation}\label{eq:BSpline-tensor-prod}
\mathbf{B}_{\mathbf{i}}^\mathbf{p}\left(\boldsymbol\xi\right) = B_{i_1}^{p_1}\left(\xi_1\right) B_{i_2}^{p_2}\left(\xi_2\right) B_{i_3}^{p_3}\left(\xi_3\right),
\end{equation}
where $\mathbf{p}=\left(p_1,p_2,p_3\right)$ and $\mathbf{i}=\left(i_1,i_2,i_3\right)$ is a multi-index in the set
\begin{equation}
\mathcal{I} = \left\lbrace \left(i_1,i_2,i_3\right) : 1 \leq i_d \leq n_d\right\rbrace.
\end{equation}
We will denote by $S_{\vect{p}}(\bm{\Xi})$ the multivariate B-spline space spanned by the basis functions $\mathbf{B}_{\mathbf{i}}^\mathbf{p}$.

\begin{figure}
\centering
\includegraphics[width=0.45\textwidth]{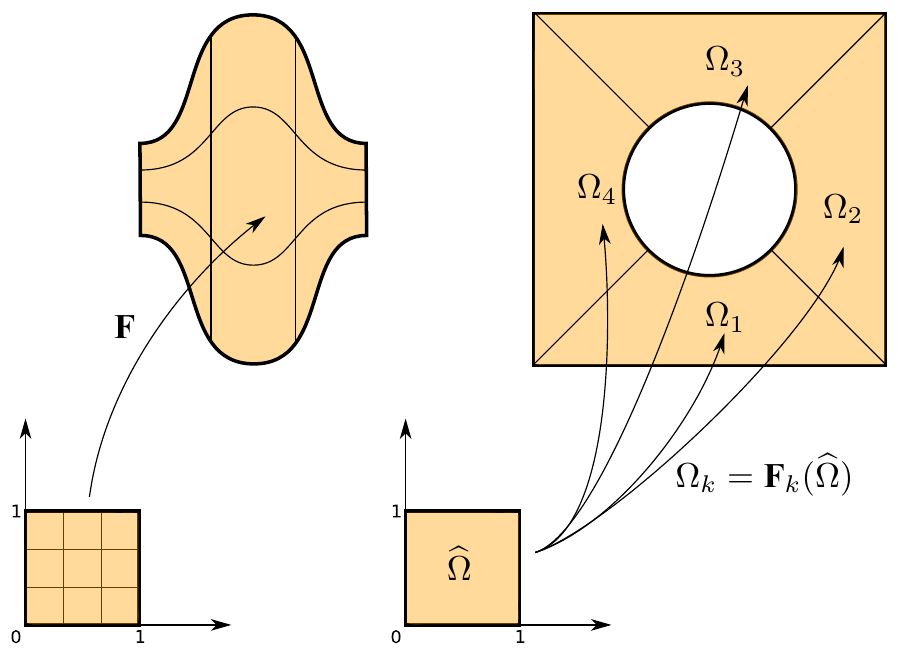}
\caption{The mesh in the physical space is given by the transformation through the NURBS mapping $\mathbf{F}$ of the knot lines in the reference domain (on the left). On the rigth an example of a multipatch mapping to represent complicated structures.\label{fig:multipatch}}
\end{figure}

To build a B-spline or \gls{nurbs} curve, we start by defining a set of \emph{control points}. These act as weights for the linear combination of the basis functions, giving the mapping to the physical space. In particular, given $n$ one-dimensional basis functions $N_i^p$ and $n$ control points $\mathbf{P}\in\mathbb{R}^d$, $i = 1, \dots, n$, a curve parameterization is given by:
\begin{equation}\label{eq:CADCurve}
\mathbf{C}\left(\xi\right)=\sum_{i=1}^{n} \mathbf{P}_{i} N_i^p\left(\xi\right).
\end{equation}
The control points define the so called control mesh but this does not, in general, conform to the actual geometry. On the contrary, the physical mesh is a decomposition of it: the mesh element edges are the image of the knot lines through the mapping~\eqref{eq:CADCurve} (see Fig. \ref{fig:multipatch} on the left).

It is straightforward to notice that the continuity of a \gls{cad} curve is controlled by the basis functions (by knot repetitions in particular), while the control points define the shape without altering the curve continuity. Moreover, as a consequence of the locality of the basis functions, moving a single control point can affect the geometry of no more than $p+1$ elements of the curve. 

The main advantage of using \gls{nurbs} over B-spline curves is the possibility to exploit both the control points and the weights in \eqref{eq:nurbs_basis} to control the local shape: as $w_{i}$ increases, the curve is pulled closer to the control point $\mathbf{P}_{i}$, and vice versa. While non-rational splines (or B\'ezier curves) can approximate a circle, they are unable to represent it exactly. Rational splines, however, overcome this issue.

In many real-world applications, the computational domain may be too complicated to be represented by a single \gls{nurbs} mapping from the reference domain to the physical space. This could be due, for example, to topological reasons or to the presence of different materials. In these cases it is common practice to resort to the so called \emph{multipatch} approach \cite{Cottrell_2009aa}. Here the physical domain is split into simpler subdomains $\Omega_k$ such that each one of them is the image of the reference space through a map $\mathbf{F}_k$ of the type given in \eqref{eq:mapping}. An example of such a situation is depicted in Fig.~\ref{fig:multipatch} on the right. 

Finally, we care to mention that the possibility to use the same space for the parameterization of geometry and solution (the so called \emph{isoparametric} approach), is particularly interesting when dealing with shape deformations, either coming from the solution of mechanical problems or from optimization or sensitivity analysis procedures. The deformation degrees of freedom can simply be added to the geometry control points to obtain the deformed domain, and the internal parameterization automatically follows with no need of cumbersome remeshing procedures.

However, we will see that in the case of Maxwell the isoparametric concept must be relaxed as the derivative of a \gls{nurbs} is not a \gls{nurbs} function. Therefore, one uses \gls{nurbs} for the mapping and B-splines as weighting and ansatz functions.

\subsection{IGA Conforming Spaces}\label{sec:iga-derham}

Analogously to classical \gls{fem}, \gls{iga} is (typically) based on a Galerkin approach: the equations are written in their variational formulation, and the solution is sought in a finite dimensional space with the correct approximation properties. In \gls{iga}, however, the basis function space is inherited from the one used to parametrize the geometry.

Let us now consider a domain $\Omega \in \mathbb{R}^d$ that can be exactly parametrized with a mapping $\mathbf{F}$ of the type in \eqref{eq:CADCurve}, i.e.
\begin{equation}\label{eq:mapping}
\mathbf{F}:\hat{\Omega}\rightarrow\Omega,
\end{equation}
with $\hat{\Omega}$ the reference domain $[0,1]^d$. We will denote by $D\mathbf{J}$ the Jacobian of the transformation.

We define the following pull-back functions
\begin{subequations}\label{eq:pull-backs}
\begin{align}
\iota_0(v) &:= v \circ \mathbf{F} && v \in \Hone{{\Omega}}{}\\
\label{eq:iota1}
\iota_1(\vect{v}) &:= \left(D\mathbf{F}\right)^\top\!\left(\vect{v} \circ \mathbf{F}\right) && \vect{v} \in \Hcurl{{\Omega}}{}\\
\label{eq:iota2}
\iota_2(\vect{v}) &:= \text{det}\left(D\mathbf{F}\right)\left(D\mathbf{F}\right)^{-1}\!\left(\vect{v} \circ \mathbf{F}\right) && \vect{v} \in \Hdiv{{\Omega}}{}\\
\iota_3(v) &:= \text{det}\left(D\mathbf{F}\right)\left(v \circ \mathbf{F}\right) && v \in \Ltwo{{\Omega}}.
\end{align}
\end{subequations}
It can be shown that \eqref{eq:iota1} and \eqref{eq:iota2} preserve the curl and the divergence, respectively, from the reference domain to the physical one \cite{Monk_2003aa,Yin_2014aa}. Due to this property, the following commuting de Rham diagram holds:
\begin{equation}\label{eq:deRham}
\hspace*{-0.3cm}\adjustbox{scale=0.85}{\begin{tikzcd}
\Hone{\hat{\Omega}}{} \arrow[r,"\hat{\nabla}"] &\Hcurl{\hat{\Omega}}{} \arrow[r,"\hat{\nabla}\times"] &\Hdiv{\hat{\Omega}}{} \arrow[r,"\hat{\nabla}\cdot"] &\Ltwo{\hat{\Omega}}\\
\Hone{\Omega}{} \arrow[r,"\nabla"] \arrow[u,"\iota_0"] &\Hcurl{\Omega}{} \arrow[r,"\nabla\times"] \arrow[u,"\iota_1"] &\Hdiv{\Omega}{} \arrow[r,"\nabla\cdot"] \arrow[u,"\iota_2"] &\Ltwo{\Omega} \arrow[u,"\iota_3"]
\end{tikzcd}}
\end{equation}
Here we have also introduced the usual Sobolev space $H^1$ of $L^2$ functions with square integrable gradient and the space $\Hdiv{\cdot}{} $ of functions with weak divergence in $L^2$. 

To obtain a conforming discretization of $\mathbf{H}(\textbf{curl},{\Omega})$, we search an analogous diagram in the discrete setting. By exploiting the fact that the derivative of a B-spline function is still a B-spline function \cite{Piegl_1997aa}, we start defining the sequence of B-spline spaces on the reference domain $\hat{\Omega}$
\begin{align}
S^0(\hat{\Omega}) &= S_{\vect{p}}(\bm{\Xi})\\
S^1(\hat{\Omega}) &= S_{p_{1}-1,p_{2},p_{3}}(\bm{\Xi})\times S_{p_{1},p_{2}-1,p_{3}}(\bm{\Xi})\label{eq:S1}\\
									&\hspace{4.3cm}\times S_{p_{1},p_{2},p_{3}-1}(\bm{\Xi})\hspace{0.8cm}\nonumber\\
S^2(\hat{\Omega}) &= S_{p_{1},p_{2}-1,p_{3}-1}(\bm{\Xi})\times S_{p_{1}-1,p_{2},p_{3}-1}(\bm{\Xi})\label{eq:S2}\\
									&\hspace{4.3cm}\times S_{p_{1}-1,p_{2}-1,p_{3}}(\bm{\Xi})\hspace{0.8cm}\nonumber\\
S^3(\hat{\Omega}) &= S_{\vect{p}-1}(\bm{\Xi}).\label{eq:L2discreteconforming}
\end{align}
It has been proven \cite{Buffa_2010aa} that, using these spaces, a discrete counterpart to \eqref{eq:deRham} can be constructed
\begin{equation}\label{eq:deRham-iga}
\begin{tikzcd}
S^{0}(\hat{\Omega}) \arrow[r,"\hat{\nabla}"] &S^{1}(\hat{\Omega}) \arrow[r,"\hat{\nabla}\times"] &S^{2}(\hat{\Omega}) \arrow[r,"\hat{\nabla}\cdot"] &S^{3}(\hat{\Omega}).
\end{tikzcd}
\end{equation}

To define the spaces in the physical domain, we use the pull-backs~\eqref{eq:pull-backs}, i.e. a conforming discretization of $\Hcurl{\Omega}{}$ is given by
\begin{equation}\label{eq:s1space}
S^1(\Omega) = \left\lbrace \vect{v} = \iota_1^{-1}(\hat{\mathbf{v}}),\hat{\vect{v}}\in S^1(\hat{\Omega}) \right\rbrace.
\end{equation}

In case of a multipatch domain, e.g. Fig.~\ref{fig:multipatch}, a global discretization space is constructed. Neighbouring patches are required to share either a full edge or a full face, i.e. no T-junctions are allowed. On each patch the matrix assembly is then performed independently and the common degrees of freedom are matched one-to-one through static condensation. It is clear that the multipatch approach reduces the global regularity to $C^0$, although inside each patch the discretization remains highly smooth. 

\section{Available Software}

For those researchers interested in \gls{iga} and wanting to test how it works in practice, a perfect way to start is GeoPDEs ~\cite{deFalco_2011aa,Vazquez_2016aa}, an open source and freely distributed package written in MATLAB/Octave language \cite{Mathworks_2009aa,Eaton_2015aa}. The GeoPDEs package was developed with a double aim. First, to serve as a didactic tool to introduce \gls{iga} to other researchers and students. For this reason the package contains a long list of examples, and all the functions include a detailed documentation accessible from MATLAB with the \texttt{help} command. Secondly, to serve as a research tool for fast prototyping and for testing new ideas and methods in \gls{iga}, and that is the main reason why the code is developed in MATLAB, which is a {\it de facto} standard for prototyping of numerical algorithms.

GeoPDEs contains all what is required for the implementation of \gls{iga}: evaluation of B-splines and \gls{nurbs} functions, matrix assembly, imposition of boundary conditions, etc. It also contains functions to export the numerical results to ParaView \cite{ParaView} for post-processing. Solving a new problem can often be easily accomplished by calling existing functions, which only require minor modifications with respect to the already existing examples. For example, most applications we present in section \ref{sec:applications} have been implemented with GeoPDEs.

One of the most interesting features is that, as far as we know, GeoPDEs is the only {MATLAB} software that contains the spline complex of Section~\ref{sec:discretisation}, which is crucial for the application of IGA in computational electromagnetism. Moreover, GeoPDEs is under constant development, and new features appear from time to time. For instance, a very recent addition is the introduction of IGA adaptive methods based on hierarchical B-splines (see~\cite{Garau_2016aa} for the details).
							   
In Listing \ref{lst:example-code}, we report a simple example of how to solve Maxwell's eigenvalue problem on a geometry given in the file \texttt{geo\_file.mat}. First we extract the knot vectors defining the geometry, we create a refinement and modify them like in \eqref{eq:S1}. Then we construct the mesh and the curl-conforming space \eqref{eq:s1space} (line 23). The functions \texttt{op\_curlu\_curlv\_tp} and \texttt{op\_u\_v\_tp} are responsible for the assembly of the curl-curl matrix $\mathbf{K}$ and the mass matrix $\mathbf{M}$ respectively, exploiting the tensor product structure. In lines 33-38 we exclude from the computation the PEC degrees of freedom at the boundary and finally we call \texttt{eig} to compute the eigenmodes.

GeoPDEs provides a compromise between clarity and efficiency. It can be applied to the solution of relatively large three-dimensional problems such as those in Section~\ref{sec:applications}. For researchers interested in even larger problems, or more efficient implementations, there are several libraries written in C or C++. Here, we care to mention, \texttt{igatools}~\cite{Pauletti_2015aa}, \texttt{G+Smo}~\cite{Juettler_2014aa} and PetIGA~\cite{Dalcin_2016aa} (which recently added curl- and div-conforming spline discretizations~\cite{Sarmiento_2016aa}). A more detailed list of available \gls{iga} software can be found in~\cite{Nguyen_2015aa}. 

\begin{lstlisting}[language=Octave,
								 caption={A simple example of how to solve Maxwell's eigenvalue problem in GeoPDEs\label{lst:example-code}},
								 basicstyle=\small\ttfamily,
								 commentstyle=\it,
								 keywordstyle=\bf,
								 numbers=left,
							   numbersep=5pt,
								 xleftmargin=15pt,
								 breaklines=true,
								 ]
% Load Geometry
geo = geo_load ('geo_file.mat');

% Define Hcurl conforming knot vectors
[knots, zeta] = kntrefine (geo.nurbs.knots, nsub, degree, regularity);
[knots_hcurl, degree_hcurl] = knt_derham (knots, degree, 'Hcurl');

% Construct the Mesh
rule = msh_gauss_nodes (nquad);
[qn, qw] = msh_set_quad_nodes (zeta, rule);
msh = msh_cartesian (zeta, qn, qw, geo);

% Construct the Space
scalar_spaces = cell (msh.ndim, 1);
for idim = 1:msh.ndim
  scalar_spaces{idim} = sp_bspline (knots_hcurl{idim}, degree_hcurl{idim}, msh);
end
space = sp_vector (scalar_spaces, msh, 'curl-preserving');

% Assemble the matrices
K = op_curlu_curlv_tp (space, space, msh);
M = op_u_v_tp (space, space, msh, @(x,y) mu0*eps0*ones (size(x)));

% Apply PEC Boundary Conditions
drchlt_dofs = [];
for iside = 1:numel (space.boundary)
  drchlt_dofs = union (drchlt_dofs, space.boundary(iside).dofs);
end
int_dofs = setdiff (1:space.ndof, drchlt_dofs);

% Solve the Eigenvalue Problem
eigv = eig (full (K(int_dofs, int_dofs)), full (M(int_dofs, int_dofs)));
\end{lstlisting}
 
\section{Applications}\label{sec:applications}

\begin{figure}
\centerline{\includegraphics[width=.7\columnwidth]{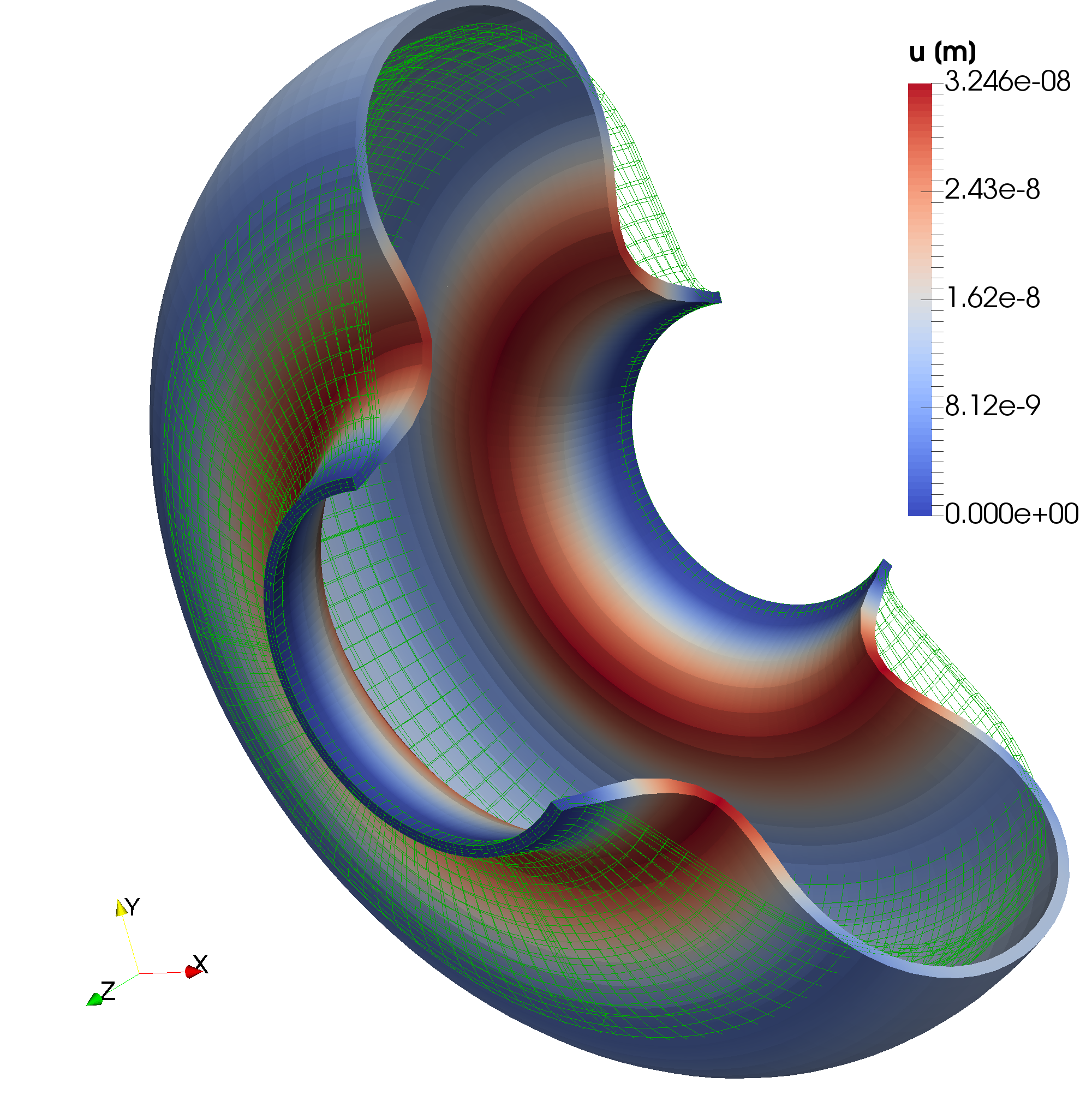}}
\caption{Lorentz detuning in a TESLA cavity cell: In green the design geometry, in color the displaced walls. The magnitude of the displacement $\mathbf{u}$ is enhanced by a factor $5 \cdot 10^5$ for visibility.}\label{fig:lor-detuning-diplacement}
\end{figure}

While \gls{iga} is already wide spread in the mechanical engineering community, real-world applications in the context of electrical engineering are still rare. This is particularly true for three-dimensional problems that require the more complicated curl- and div-conforming spline spaces from equations \eqref{eq:S1} and \eqref{eq:S2}. In \cite{Manh_2014aa} the two-dimensional shape optimization of a magnetic density separator was proposed, and in \cite{Lee_2016aa} the optimization of ferromagnetic materials in a magnetic actuator was shown. Recently, researchers have also started to investigate Isogeometric Analysis of integral equations in the context of electrical engineering, e.g.~\cite{Sekine_2014aa,Sekine_2014ab,Li_2015aa,Simpson_2013aa}. 

In the following subsections the application of \gls{iga} is demonstrated for several real-world examples in two and three dimensions and an outlook on isogeometric \glspl{bem} is given. The aim is not to give a precise description but rather to illustrate that \gls{iga} is indeed a useful tool for the Compumag community.   
\subsection{Radio Frequency Cavities}\label{sec:detuning}

To achieve acceleration of the particle bunches in particle accelerator \gls{rf} cavities, the electromagnetic field has to oscillate at a very specific frequency, synchronously to the movement of the charges. The eigenfrequency is determined by the shape of the cavity walls, which is therefore critical for the design of any cavity. However, the high-energy field exerts a radiation pressure on the walls, which impresses a mechanical deformation of the domain. Albeit small, this deformation may lead to a significant shift of the resonance frequency. This effect is known as \emph{Lorentz detuning}~\cite{Devanz_2002aa,Gassot_2002aa,Zaplatin_2006aa} and needs to be predicted with high precision in order to achieve a robust cavity design. 

Applying standard \glspl{fem} present two main problems: the domain boundary is approximated by polynomials and the deformation of the cavity walls may require an interpolation and remeshing step or an ad-hoc mesh movement procedure. In \cite{Schreiber_2006aa} the MpCCI Multiphysics Interface \cite{MpCCI} was used for exchanging geometry and solution between the CST Microwave predecessor MAFIA based on the Finite Integration Technique and the software package ParaFep based on Finite Elements \cite{CST,Niekamp_2002aa}. \gls{iga} is able to overcome these issues allowing an exact representation of the geometry, leading to higher accuracy, and a direct application of the computed deformation to the design shape, without any further approximation. Finally it offers the possibility to obtain highly smooth solutions, which can prove extremely valuable for particle tracking applications \cite{Gjonaj_2009aa}.

\paragraph{Maxwell's Eigenvalue Problem.} As a first example we consider the academic case of the computation of the eigenmodes in a cylindrical resonating cavity (\emph{pill-box}), where the eigenmodes can be computed analytically. The fields are assumed to be time-harmonic and oscillating in vacuum ($\epsilon=\epsilon_0$, $\mu=\mu_0$), with no charges or currents. The first order system \eqref{eq:Maxwell} can then be rewritten as a second order equation for the electric field $\vect{E}$ only
\begin{equation}\label{eq:curl-curl}
\curl{\curl{\vect{E}}} = \mu_0 \epsilon_0 \omega^2 \vect{E}.
\end{equation}
The solution of problem \eqref{eq:curl-curl} is a sequence of eigenmodes $(\omega_m^2, \vect{E}_m)$ which represents the excitable modes in the cavity. The quantities $\omega_m$ are the resonance frequencies $f_m = \omega_m / 2\pi$.

\begin{figure}
\begin{minipage}{.48\columnwidth}
\includegraphics[width=\textwidth]{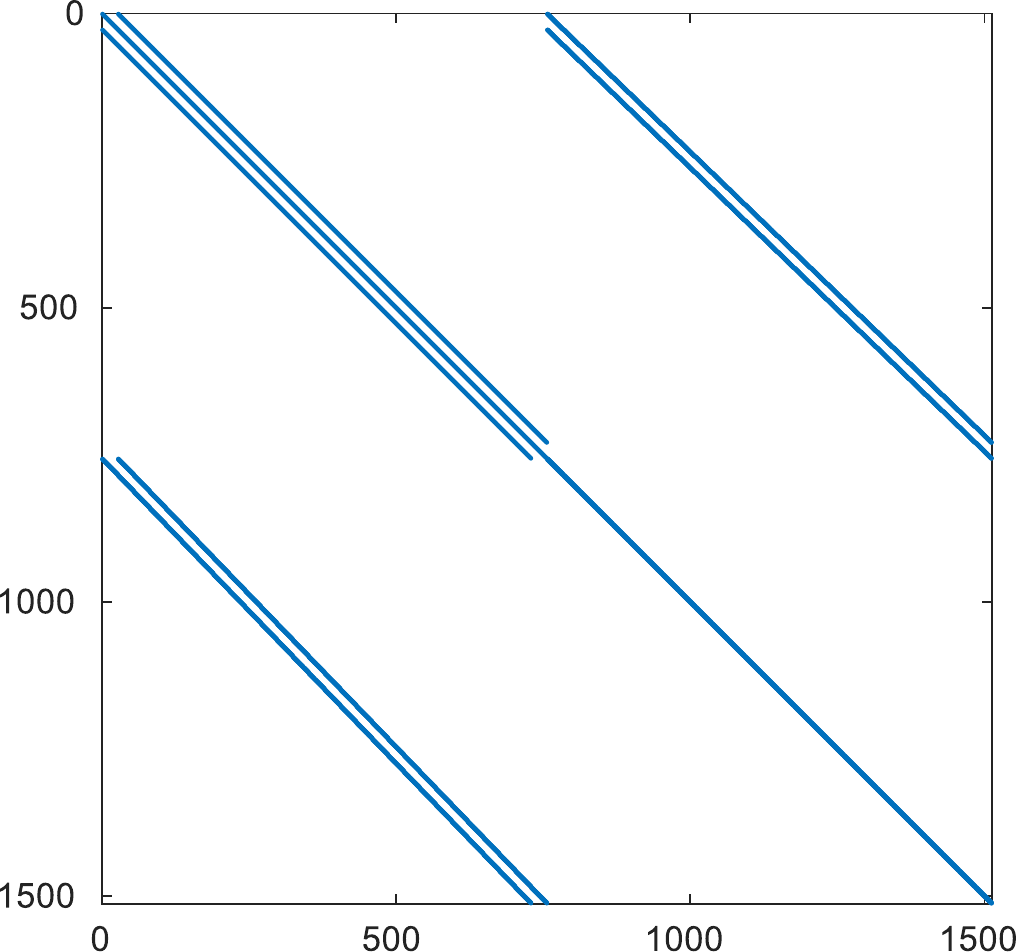}
\end{minipage}\hfill
\begin{minipage}{.48\columnwidth}
\includegraphics[width=\textwidth]{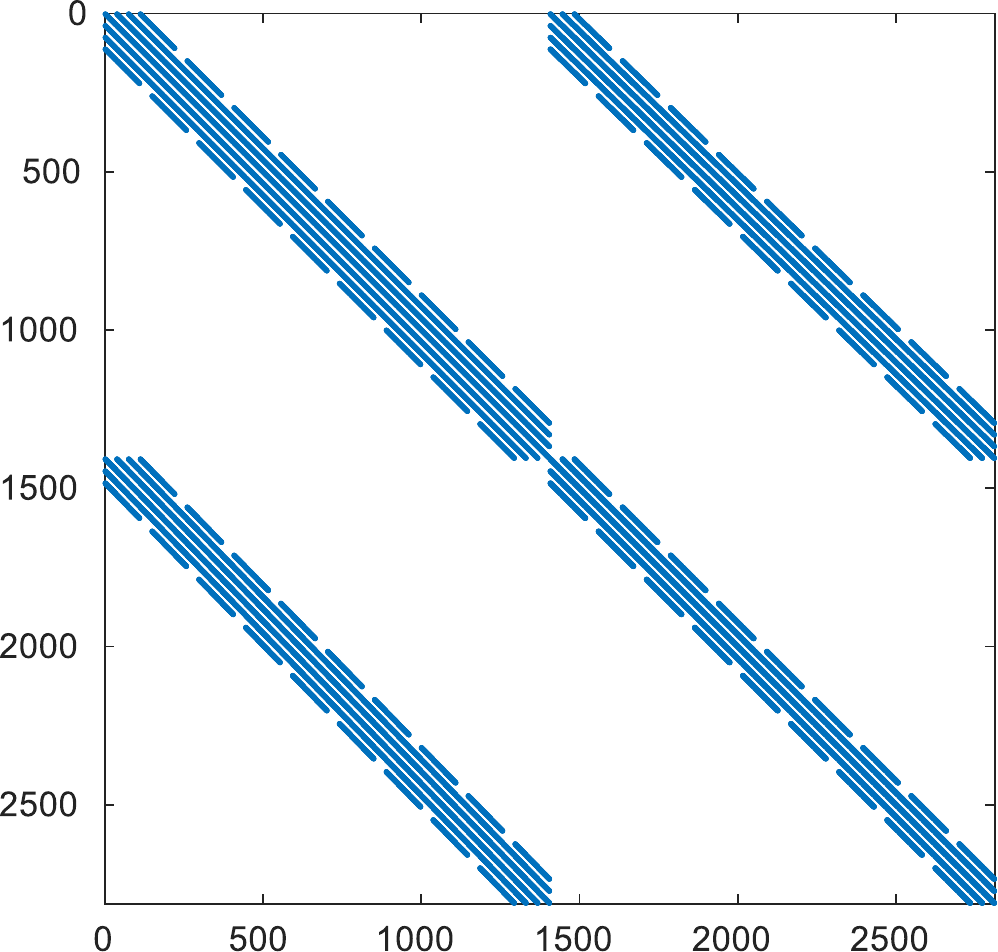}
\end{minipage}
\caption{Spyplots of mass matrices for different degrees. On the left the \gls{fem} lowest order case (in this case \gls{iga} and \gls{fem} coincide). The ratio between non-zero elements and total number of elements in the matrix is \num{0.0045}. On the right $C^2$ B-spline basis functions of degree $p=3$. In this case the ratio is \num{0.0199}.\label{fig:spyplots}}
\end{figure}

\begin{figure}
\includegraphics[width=0.45\textwidth]{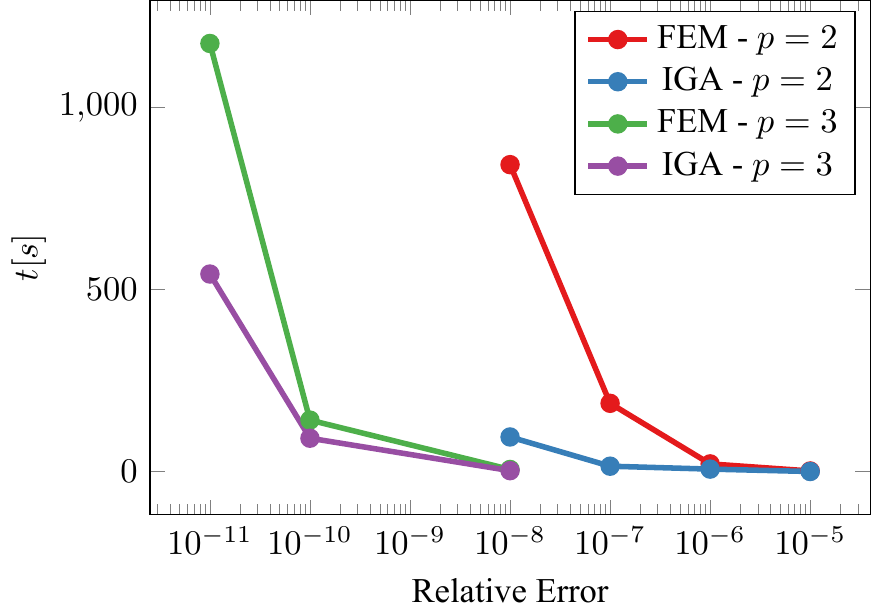}
\vspace{-1.5em}
\caption{Computational time required to solve the eigenvalue problem with ARPACK for finding the first accelerating mode in the pill-box cavity, with a prescribed accuracy. The IGA implementation was performed in GeoPDEs while for the FEM simulation CST
EM STUDIO \cite{CST} was used.\label{fig:pillbox-time}}
\end{figure}

Equation \eqref{eq:curl-curl} can be discretized using B-splines belonging to the space \eqref{eq:s1space} in order to obtain a generalized eigenvalue problem
\begin{equation}
\mat{K}\vect{e} = \omega^2\mat{M}\vect{e},
\end{equation}
with $\mat{K}$ and $\mat{M}$ the curl-curl and mass matrix respectively, and $\vect{e}$ the vector containing the electric field degrees of freedom. For the software implementation one can follow the lines of Listing~\ref{lst:example-code}. The matrices obtained with the \gls{iga} discretization typically present a larger bandwidth compared to their \gls{fem} counterparts (see Fig.~\ref{fig:spyplots}), but, as previously mentioned, they are also smaller for a given accuracy. We present a comparison between the two approaches in terms of efficiency of the solution of the eigenvalue problem. A set of matrices, obtained from meshes with increasing refinement, was generated for order 2 and 3 polynomial \gls{fem} basis functions using the proprietary software CST \cite{CST}, and exported to MATLAB. The same Arnoldi/Lanczos solver is used for the IGA matrices and for the FEM ones. In Fig.~\ref{fig:pillbox-time} we report the computational time necessary for attaining a prescribed level of accuracy. Given the higher accuracy-per-degree-of-freedom of \gls{iga}, it is possible to achieve a considerable speed-up \cite{Corno_2016aa}.

\paragraph{Lorentz Detuning.} The procedure we propose is straightforward. First, we compute the fields in the cavity by solving \eqref{eq:curl-curl}. As a second step we solve the mechanical problem for the cavity walls. Given the small deformations involved we use the linear elasticity model
\begin{equation}
	\div{\left(2\eta\nabla^{\left(S\right)}\vect{u}+\lambda\mathbf{I}\nabla\cdot\vect{u}\right)} = p\vect{n} \label{eq:mech}
\end{equation}
for the deformation $\vect{u}$, with $\eta$, $\lambda$ the Lamé constants of niobium and $p\vect{n}$ the pressure applied. The symbol $\nabla^{\left(S\right)}$ denotes the symmetric gradient operator. The electromagnetic problem \eqref{eq:curl-curl} couples into the mechanical problem by the radiation pressure
\begin{align}
	p_{\text{rad}} = 
	&-\frac{1}{4}\epsilon_0\Big(\vect{E}_{\text{pk}}\cdot\vect{n}_c\Big)\left(\vect{E}_{\text{pk}}^{*}\cdot\vect{n}_c\right)\\
	&+ \frac{1}{4}\mu_0\Big(\vect{H}_{\text{pk}}\times\vect{n}_c\Big)\cdot\left(\vect{H}^{*}_{\text{pk}}\times\vect{n}_c\right)\nonumber
\end{align}
where $\vect{E}_{\text{pk}}$ and $\vect{H}_{\text{pk}}$ are the field peak values, $\vect{n}_c$ is the outside normal to the cavity and $^*$ denotes the complex conjugate operator.

As mentioned above the computed deformation $\vect{u}$ can be directly applied to the control polygon of the cavity geometry to obtain the deformed cavity. The solution of Maxwell's eigenvalue problem on this domain gives the frequency shift.

For the computation of the frequency shift in the case of a pill-box cavity and a comparison with a sensitivity analysis procedure we refer the interested reader to \cite{Corno_2016aa}.

In Fig.~\ref{fig:lor-detuning-diplacement} we depict the deformation for the more realistic case of a \SI{1.3}{\giga\hertz} TESLA cavity \cite{Aune_2000aa}. This deformation is typically of the order of tenths of \si{\nano\metre}, nevertheless, this leads to a measurable frequency shift in the order of hundredths of \si{\hertz} that needs to be addressed during operation. The computed shift is approximately \SI{1}{\kilo\hertz} \cite{Corno_2017aa} which is in good agreement with the literature.

\begin{figure}
\centering
\includegraphics[width=0.45\textwidth]{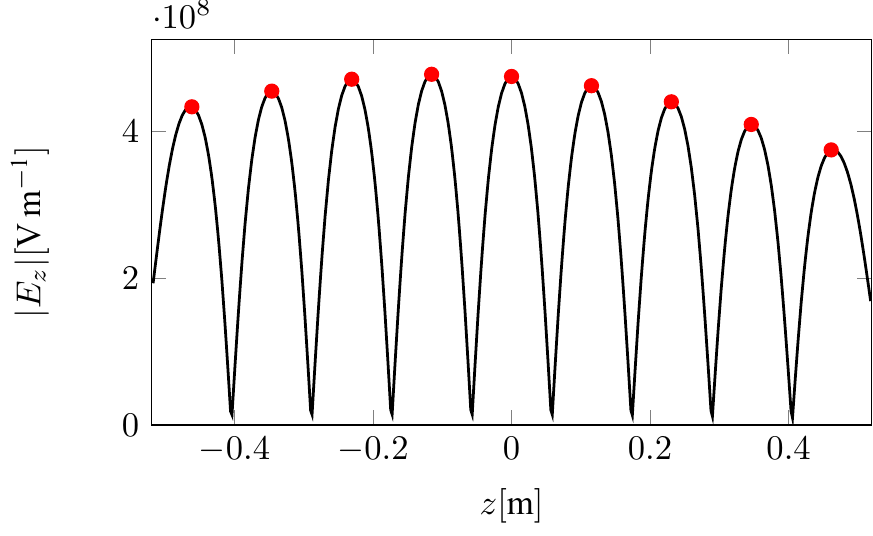}\\
\includegraphics[width=0.45\textwidth]{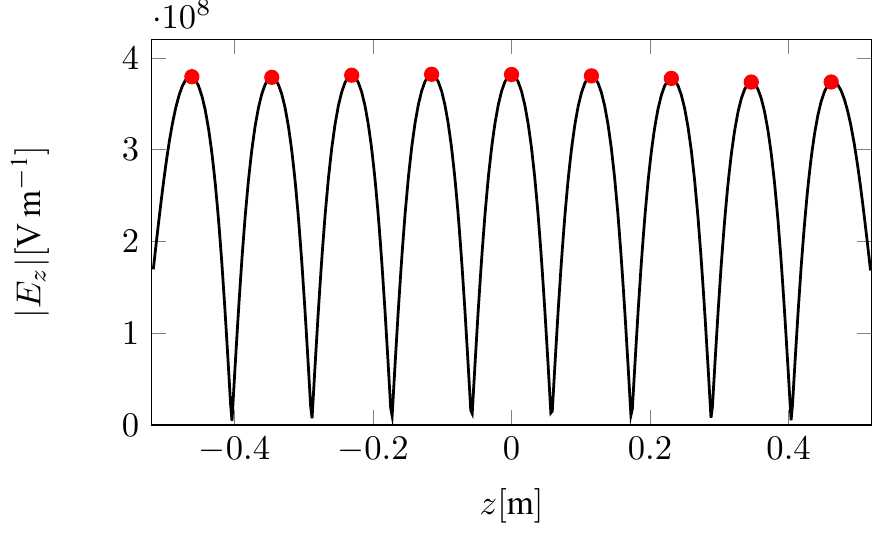}
\vspace{-1.5em}
\caption{Absolute value of the longitudinal electric field $E_z$ in the untuned (top) and tuned (bottom) TESLA cavity. The computed value for the field flatness improve from $\eta_1 = \SI{76.81}{\percent}$ and $\eta_2 = \SI{92.39}{\percent}$ to  $\eta_1 = \SI{97.76}{\percent}$ and $\eta_2 = \SI{99.15}{\percent}$.\label{fig:field-flatness}}
\end{figure}
\paragraph{Field Flatness Tuning.} When dealing with multi-cell cavities such as the TESLA design, small variations between the cells' shape are sufficient to substantially alter the
field profile. Of particular interest for the correct operation of \gls{rf} structures is to achieve a uniform energy distribution in each cell. This presents two advantages: it maximizes the accelerating voltage of the cavity (i.e. the net energy gained by the particles) and it minimizes the peak surface fields, which are responsible for electric field emission or quench in the superconducting walls \cite{Padamsee_2008aa}.

We set $z$ as the longitudinal axis of the cavity and we denote by $E_{\text{pk},j}$ the peak value of $E_z(r=0,z)$ in the $j$-th cell. The field flatness is typically measured by two quantities:
\begin{subequations}\label{eq:field-flatness-eta}
\begin{align}
\eta_1 &= \frac{1-\left(\max_j\left|E_{\text{pk},j}\right|-\min_j\left|E_{\text{pk},j}\right|\right)}{\mathbb{E}\left(\left|E_{\text{pk},j}\right|\right)}\\
\eta_2 &= 1 - \frac{\text{std}\left(E_{\text{pk},j}\right)}{\mathbb{E}\left(\left|E_{\text{pk},j}\right|\right)},
\end{align}
\end{subequations}
where $\mathbb{E}$ and std denote the expected value and the standard deviation operators. Both $\eta_1$ and $\eta_2$ are typically required to be $\geq 0.95$ for a well tuned cavity.

In practice, the tuning is performed through mechanical deformation. The field flatness is measured and, using a circuit model for the cells as capacitively coupled LC oscillators, the required frequency shift for each cell is computed \cite{Padamsee_2008aa}. To lower (resp. increase) the frequency, each cell is shortened (elongated) along the $z$ axis by a tuning machine which clamps the cavity between cells and applies forces to obtain a permanent deformation \cite{Kreps_1996aa}.

Numerically, the tuning is performed through a multi-objective shape optimization procedure that aims at maximizing the field flatness parameters \eqref{eq:field-flatness-eta} and achieving a resonance frequency of \SI{1.3}{\giga\hertz}. The geometry parameters that are typically chosen for the optimization are the length of the two end-cups half-cells and the equatorial radius of the cavity. The first two parameters strongly influence the field flatness, while the equatorial radius is mainly responsible for fixing the frequency of the cavity. The three parameters are used for a non-linear constrained optimization procedure using the SQP method. The bounds for the parameters are set to $\pm \SI{1.5}{\milli\metre}$.

In Fig.~\ref{fig:field-flatness} the longitudinal electric field along the cavity axis before and after the tuning is depicted.

The advantage of \gls{iga} when dealing with shape optimization is the possibility to affect both the geometry and the mesh at the same time by simply moving the control points. For each configuration there is no need of a remeshing step that might introduce undesired noise in the computation. Furthermore, \gls{nurbs} geometries are well suited for the computation of shape derivatives in order to obtain gradient information.

\subsection{Electric Machine Simulation}\label{sec:pmsm}

For the modeling of electric machines, a common approach is the magnetostatic formulation where the eddy currents and the displacement currents are neglected, see e.g. \cite{Salon_1995aa}. Under these assumptions, by introducing the magnetic vector potential $\vect{A}$, with $\vect{B} = \curl{\vect{A}}$, one obtains
\begin{equation}
  \nabla \times (\nu \nabla \times \vect{A}) = \vect{J}
\end{equation}
on the domain $\Omega$ with appropriate boundary and gauging conditions. Furthermore, given the motor structure, it is often sufficient to solve the problem only on a 2D cross section. One obtains a Poisson problem for the longitudinal component $\vect{A}=[0,0,u]^\top$
\begin{equation}\label{eq:machine-poisson}
- \div(\nu \grad u) = J_z,
\end{equation}
where $\nu$ is the space-dependent reluctivity, and $J_z$ represents the current excitations due to the presence of coils and/or permanent magnets.

\begin{figure}
	\centering
	\includegraphics[width=0.4\textwidth]{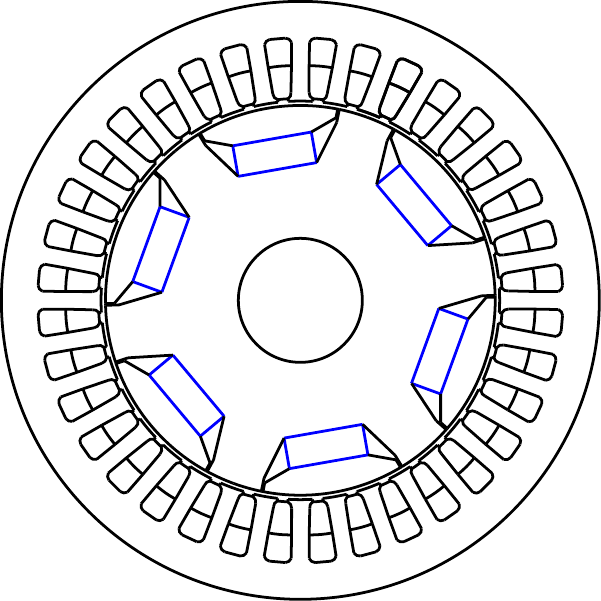}
	\caption{Computational domain $\Omega$ of a \gls{pmsm} model as discussed in \cite{Bontinck_2017aa,Bhat_2017ab}.\label{fig:pmsm-geo}}
\end{figure}

In Fig.~\ref{fig:pmsm-geo} the geometry of a \gls{pmsm} is depicted. Although only two dimensional, the topology and the presence of different materials requires the construction of a high number of patches (12 for the rotor, 78 for the stator), which is still low compared to the number of elements in a \gls{fem} discretization. In \cite{Bontinck_2017aa} we propose a harmonic stator-rotor coupling method to overcome this issue and to allow for easy handling of the machine rotation.

The application of \gls{iga} to machine simulation is particularly interesting for the possibility to exactly represent its circular shape and, even more so, for the smoothness of the computed fields. The evaluation of torques and electromotive force (EMF) are often calculated from the fields in the air gap using the Maxwell's stress tensor. Since the results obtained through this approach are very sensitive to the representation and to the discretization of the air gap \cite{Howe_1992aa}, the higher continuity of the isogeometric solutions has been proved beneficial.

\begin{figure}
\centering
\includegraphics[width=0.45\textwidth]{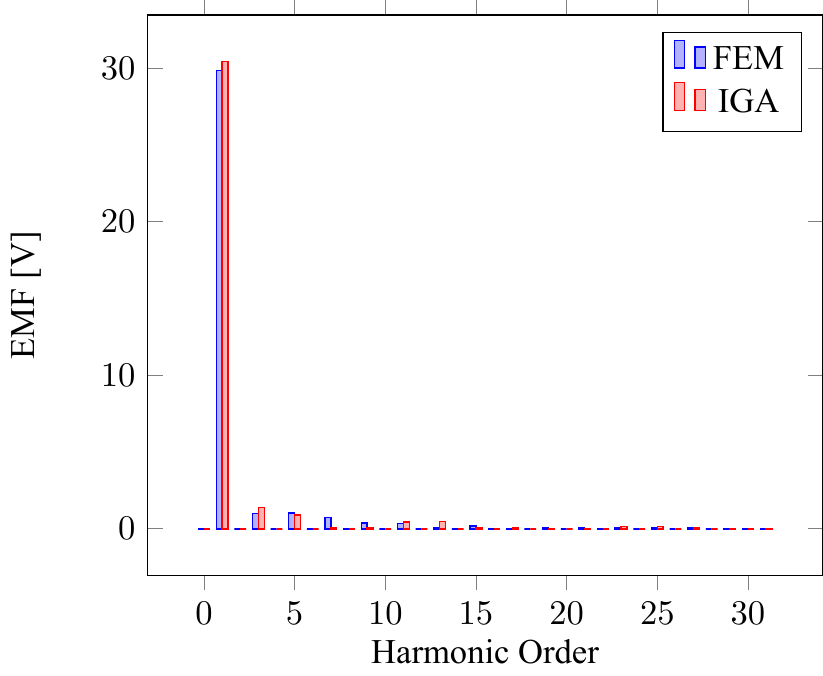}
\caption{Spectrum of the first 32 modes of the EMF of the PMSM, cf. \cite{Bontinck_2017aa,Bhat_2017ab}. \label{fig:bar}}
\label{fig:spectra}
\end{figure}

In Fig.~\ref{fig:bar} the spectrum of the first 32 harmonics of the EMF for the \gls{pmsm} is depicted. The results show good agreement with the ones obtained with a classical \gls{fem} simulation, but the \gls{iga} system is considerably smaller \cite{Bontinck_2017aa,Bhat_2017ab}.

One further interesting possibility for electric machine simulation taking into account the machine rotation would be a coupling strategy using classical \gls{iga} on the stator and the rotor and an IGA-BEM discretization in the air gap region (see e.g. \cite{Kurz_1998aa}). In sub-section~\ref{sec:bem} we briefly introduce the \gls{bem} setting and show that it can be readily applied in conjunction with the \gls{iga} concept. 
\subsection{Accelerator Magnets}

\begin{figure}[t!]
  \vspace{-1cm}
  \centering
  \includegraphics[width=0.23\textwidth]{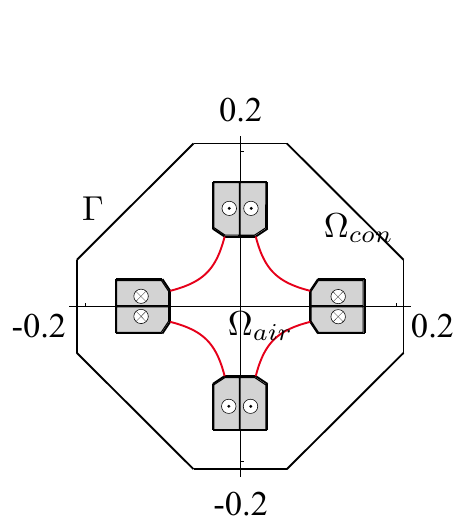}
  \includegraphics[width=0.23\textwidth]{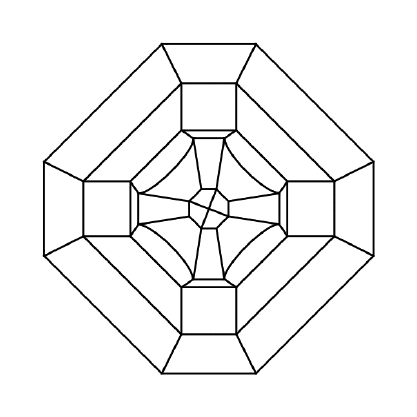}
  \caption{Quadrupole as given in Figure 7.1 of \cite{russenschuck2011field} on the left. The red pole tips are modeled as interfaces subject to uncertainty. On the right, discretization in terms of multipatch \gls{nurbs}. All units in meter. Image based on \cite{Romer_2015ab}.}
\label{fig:quadrupole}
\end{figure}

In particle accelerators normal and superconducting magnets are used for focusing and bending the particle beams. The simulation of these devices is an important task during the computer aided design process as these devices are operated at their physical limits but also as to deliver highly accurate field maps for beam dynamics simulations and, finally, the simulation of quench protection \cite{Roggen_2011aa,Cortes-Garcia_2017ab}.

\paragraph{Uncertainty Quantification.} The first magnet example is devoted to illustrate the treatment of shape uncertainties, see e.g. \cite{Clenet_2013aa}, in the context of magnet design. On the left side of Fig.~\ref{fig:quadrupole} the two-dimensional geometry of a model quadrupole magnet adopted from Fig. 7.1 of \cite{russenschuck2011field} is depicted, where the pole tips are assumed to be affected by uncertainty, e.g. due to manufacturing imperfections. The initial pole shape is described by hyperbolas, i.e., through the relation $x/a^2-y/b^2 = 1$ in local coordinates, where for the initial shape we have set $a=0.05$ and $b=0.056$, respectively. In a stochastic setting, uncertain interfaces can be modeled by choosing $a$ or $b$ as random variables, or equivalently by using any other CAD standard shape representation with random parameters. Then, a truncated Karhunen-Lo\`{e}ve expansion can be used to obtain a reduced number of uncorrelated random inputs \cite{Xiu_2010aa}. A multi-patch configuration, consisting of $36$ patches is chosen to represent the geometry as depicted in Fig. \ref{fig:quadrupole} on the right side. Each patch is described by a mapping of the type \eqref{eq:mapping} using second degree \gls{nurbs} basis functions with $\mathcal{C}^1$-continuity. In particular, this allows for an exact representation of hyperbolic shapes. 
Focusing on shape variations, we model the material to be linear with a constant permeability $\mu = 4 \pi \SI{e-2}{\henry\per\metre}$. A total piecewise constant current of $\SI{15}{\mega\ampere}$ is supplied for each of the four conductor parts of Fig. \ref{fig:quadrupole}. Homogeneous Dirichlet boundary conditions are applied on the whole boundary $\Gamma$. Following the iso-parametric concept, a basis for the discrete subspace of $H^1$ is constructed in the same space as the mapping $\mathbf{F}$. The multipole coefficients are evaluated at a reference radius of $r_0 = \SI{20}{\milli\metre}$.

We focus on the uncertainty in the quadrupole gradient, defined as $g:= 2B_2/r_0^2$ where $B_n$ is the $n$-th normal multipole coefficient. 
As we are concerned with several parameters and only one cost function, adjoint techniques are well suited to this end \cite{Hinze_2008aa,Romer_2015ab}. Here, no assumptions, despite the $\mathcal{C}^1$-smoothness, are made for the shape uncertainty and, therefore, we resort to a worst-case analysis. The maximum deviation of $g$ from its design value $g_0 = \SI{19.73}{\tesla\per\metre\squared}$ is investigated for different levels of shape parametrization, characterized by the number of free control points. On the coarsest level the hyperbola is described by three control points, however, the end points are kept fixed to avoid variability in the singular points, as this would require a more general shape calculus as presented here. Hence, only one control point per pole is subject to uncertainty. Through mesh refinement this number is increased by one on each level, up to level four. After refining the geometry, in each direction every mesh cell is divided by twenty. For an evaluation of the accuracy of the numerical approximation of the quadrupole gradient $g$, see Table \ref{tab:quadrupole_conv}. There, the error on each level is estimated with respect to a fine discretization consisting of \num{367641} total DOF, denoted as $\Delta_h g$. Additionally, the numerical computation of the shape gradient is verified. To this end in Table \ref{tab:quadrupole_conv} the maximum deviation with respect to a finite difference gradient computation, denoted $\Delta_{\mathrm{FD}}$, is given. Both estimated errors are found to be sufficiently small.
\begin{table}
\caption{Quadrupole convergence at different mesh levels. Error estimate of the shape gradient and finite difference approximation for different number of perturbed control points. The reference for $g$ is computed with \num{367641} DOF.}
\label{tab:quadrupole_conv}
\centering
\vspace{-0.2em}
\begin{tabular}{c|c|c|c|c}
\hline
Parameter & level $1$  & level $2$ & level $3$ & level $4$\\
\hline
$\Delta_h g \ /g_0$ &
\scriptsize $0.56 \%$ &
\scriptsize $0.05 \%$ &
\scriptsize $<0.01 \%$ &
\scriptsize $<0.01 \%$  \\
$\Delta_{\mathrm{FD}} \delta g$ &
\scriptsize \num{5.85e-5} &
\scriptsize \num{3.96e-5} &
\scriptsize \num{7.29e-5} &
\scriptsize \num{1.38e-4}\\
\hline
\end{tabular}
\end{table}
We emphasize that no correlation is imposed here. If knowledge of the shape perturbations were available, e.g., in terms of measurements, the correlation structure could be incorporated by means of convex constraints in a worst-case scenario context as outlined in \cite[p.10]{babuvska2005worst}. In Table \ref{tab:quadrupole_fd} numerical results for the different parametrization levels are presented. Not surprisingly, a significantly smaller worst-case estimate is obtained for the coarse parametrization. In this case, large perturbations in the control point are necessary to obtain a comparable shape perturbation to the finer parametrization levels. We compute the worst-case scenario by a Taylor expansion $\mathrm{wcs}_L(g)$ and by directly solving the optimization problem, denoted $\mathrm{wcs}^*(g)$. Here, for the latter case MATLAB's \texttt{fmincon} routine is used to carry out sequential quadratic programming. We observe a difference of about $15 \%$, and infer that the problem is rather sensitive to shape perturbations and first order approximations should be used to obtain rough estimates of the output uncertainties, solely.
\begin{table}
\caption{Quadrupole error by linearization and error estimated by means of MATLAB's \texttt{fmincon}. $\mathrm{wcs}_L(g)$ indicates the worst-case scenario computed by Taylor expansion, $\mathrm{wcs}^*(g)$ the one obtained by solving the optimization problem. Perturbation magnitude of $s = 0.1$.}
\label{tab:quadrupole_fd}
\centering
\vspace{-0.2em}
\begin{tabular}{c|c|c|c}
\hline
num. free CP & $\mathrm{wcs}_L(g)$ / $g_0$ & $\mathrm{wcs}^*(g)$ / $g_0$ & rel. diff.\\
\hline
$4$ &$7.82$  &$8.48$  & $7.78\%$ \\
$8$ &$16.77$ &$19.77$ & $15.17 \%$ \\
$12$ &$18.57$ &$21.77$ & $14.70 \%$ \\
\hline
\end{tabular}
\end{table} 

\begin{figure}
  \centering
  \includegraphics[width=0.75\columnwidth]{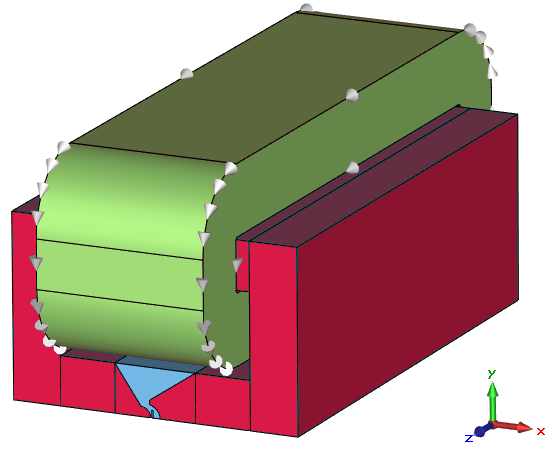}
  \caption{3D model of one half of the Stern-Gerlach magnet with Rabi-type pole tips (modeled with CST EM STUDIO \cite{CST}), see \cite{Pels_2015aa}.}
  \label{fig:CSTSternGerlachMagnet}
\end{figure}

\paragraph{Shape Optimization.} Another application in which \gls{iga} has proven advantageous is the optimization of a Stern-Gerlach magnet \cite{Pels_2015aa}. 
A Stern-Gerlach magnet (see Fig.~\ref{fig:CSTSternGerlachMagnet}) is used to magnetically separate a beam of atoms or atom clusters in order to experimentally determine their angular momentum and its spatial quantization. For this purpose, the atoms are accelerated and shot through the aperture of the magnet. To deflect the particles, the magnet should provide both, a high magnetic field strength to cause a precession of the magnetic dipoles, and a high magnetic field gradient to deflect the atoms. To obtain an acceptable resolution, an additional requirement is a high homogeneity of the magnetic field gradient in the beam area. These quantities are strongly influenced by the geometry of the pole tips. In this example, Rabi-type pole tips are considered. The optimization process consists of finding an optimal geometry for these pole tips to satisfy the before-mentioned requirements. 

The Stern-Gerlach magnet is operated with a DC current. Thus a 3D non-linear magnetostatic formulation of the Maxwell's equations is suited to calculate the magnetic field
\begin{equation}
  \nabla \times (\nu(\vect{B}) \nabla \times \vect{A}) = \vect{J},
\end{equation}
where $\nu(\vect{B})$ is the non-linear reluctivity and $\vect{J}$ the exciting current density.
Due to symmetry only one half of the magnet is considered in the optimization process, as depicted in Fig.~\ref{fig:CSTSternGerlachMagnet}. To obtain an efficient simulation, the magnet is modeled by using a combination of a magnetic equivalent circuit and a field model discretized by \gls{iga}:~the outer yoke and coils are modeled by the magnetic equivalent circuit which is extracted from a full 3D simulation. 

\begin{figure}
  \centering
  \includegraphics[width=0.8\columnwidth]{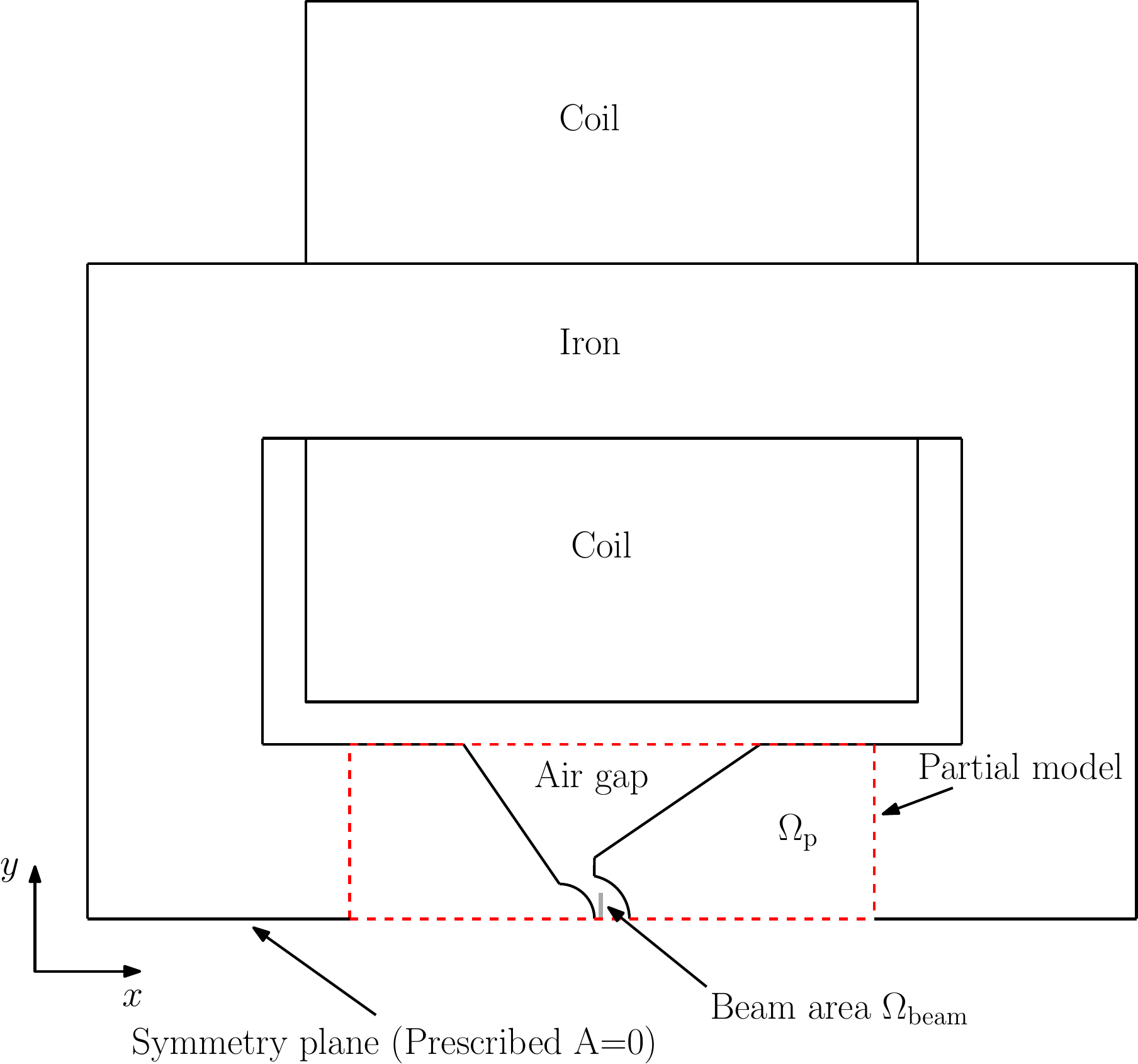}
  \caption{2D schematic of one half of the Stern-Gerlach magnet with Rabi-type pole tips, see \cite{Pels_2015aa}.}
  \label{fig:2DSternGerlachMagnet}
\end{figure}

\begin{figure}
  \centering
  \includegraphics[width=0.45\textwidth]{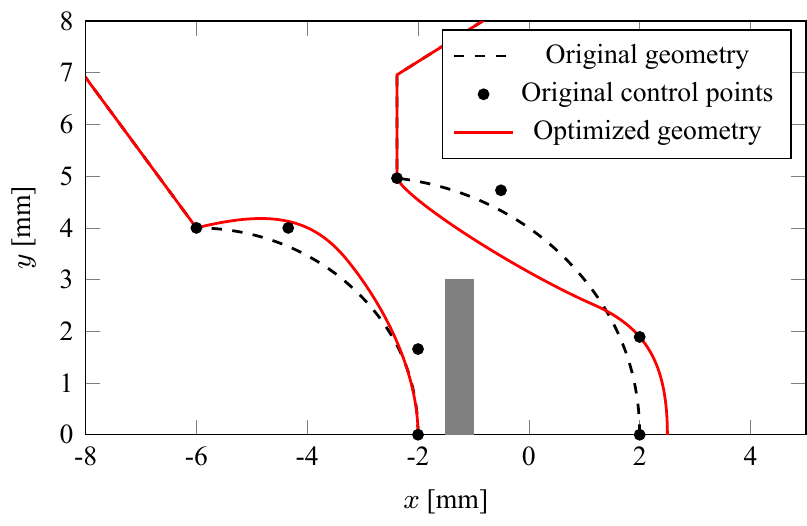}
  \caption{Optimized and original geometry of the pole tips including beam area (gray) and labeled control points of original geometry, see also \cite{Pels_2015aa}.}
  \label{fig:geometrySternGerlachMagnet}
\end{figure}

Finally, the area of the pole tips is discretized by a 2D \gls{iga} model, i.e. equation \eqref{eq:machine-poisson} on the domain $\Omega_\mathrm{p}$ as shown in Fig.~\ref{fig:2DSternGerlachMagnet}. This enables a smooth representation of the pole shapes by using NURBS and simplifies the optimization process, as the deformation of the poles is easily achieved by shifting control points and adjusting weights. The control points used for the NURBS representation of the poles are depicted in Fig.~\ref{fig:geometrySternGerlachMagnet}. The partial model is divided into 3 patches, namely the left pole region, gap and right pole region. Both the magnetic equivalent circuit and the partial model are connected by a field-circuit coupling.    
To avoid a non-linear evaluation of this coupled model and thus save computational effort, a further simplification is employed:~the permeability in the \gls{iga} partial model is frozen. This is possible as the variations in the permeability distribution due to changes in the geometry are small.

For the optimization, the quantities of interest in the beam area $\Omega_\mathrm{beam}$ of the magnet are the average magnetic field gradient and the inhomogeneity which are given by
\begin{equation} \label{equ:AvgGradient}
  \tau_\mathrm{av}=\frac{1}{|\Omega_\mathrm{beam}|}\int\limits_{\Omega_\mathrm{beam}}\tau(x,y)~\mathrm{d}\Omega
\end{equation}
and
\begin{equation} \label{equ:Homogeneity}
  \epsilon=\sqrt{\frac{1}{|\Omega_\mathrm{beam}|}\int\limits_{\Omega_\mathrm{beam}}\left(\frac{\tau(x,y)}{\tau_\mathrm{av}}-1\right)^{2}~\mathrm{d}\Omega} \;,
\end{equation}
where $\tau(x,y)=\frac{\mathrm{d} \left|\vect{B}\right|}{\mathrm{d} x}$ is the magnetic field gradient.
The goal function used for the optimization process is
\begin{equation}
  f(\mathbf{x,y,w})=\frac{\tau_{\mathrm{w}}}{|\tau_{\mathrm{av}}|}+\epsilon-\frac{\tau_{\mathrm{w}}}{|\tau_{\mathrm{av}}|}\epsilon,
  \label{eqn:goalFunction}
\end{equation}
where $\tau_{\mathrm{w}}$ is a free parameter, and $\tau_{\mathrm{av}}$, $\epsilon$ are the quantities of interest defined above. As the requirements of high average magnetic field gradient and homogeneity are, to a certain degree, antithetic, the free parameter $\tau_{\mathrm{w}}$ controls the preference of one over the other quantity. 
The overall optimization problem can be written as
\begin{equation}
  \min_{\mathbf{x,y,w}} f(\mathbf{x,y,w}),
\end{equation}
where $\mathbf{x,y,w}$ are the coordinates and weights of the control points defining the poles. These quantities are subject to certain geometrical limits to ensure the validity of the geometry.

For the simulation, the model is implemented using GeoPDEs. B-splines of order 5 are used to represent the magnetic vector potential $\vect{A}$. This way, the magnetic flux density $\vect{B}$ and even the average magnetic field gradient $\tau_\mathrm{av}$ are smooth functions opposed to conventional FE simulations. 

The original and optimized geometry of the pole shoes is depicted in Fig.~\ref{fig:geometrySternGerlachMagnet}.
The improvements of the average magnetic field gradient and homogeneity after the optimization are shown in Table \ref{tab:results}. For validation, the optimized geometry is imported into a 3D model of the magnet in the commercial software CST EM STUDIO \cite{CST}. Note that the presented optimization approach using \gls{iga} and field-circuit coupling is considerably faster than the commercial software, while offering results which are in very good agreement \cite{Pels_2015aa}.

\subsection{IGA and the Boundary Element Method}\label{sec:bem}

The framework of Isogeometric Analysis is not limited to FEM.
In recent years, Isogeometric Boundary Element Methods (BEM) gained a lot of attraction in research. 
The idea of \gls{bem} is the representation of the solution $u$ of a PDE through a representation formula, which expresses the solution via functions on the boundary of the problem domain only. 

\begin{figure}
  \centering
  \includegraphics[width=.8\columnwidth]{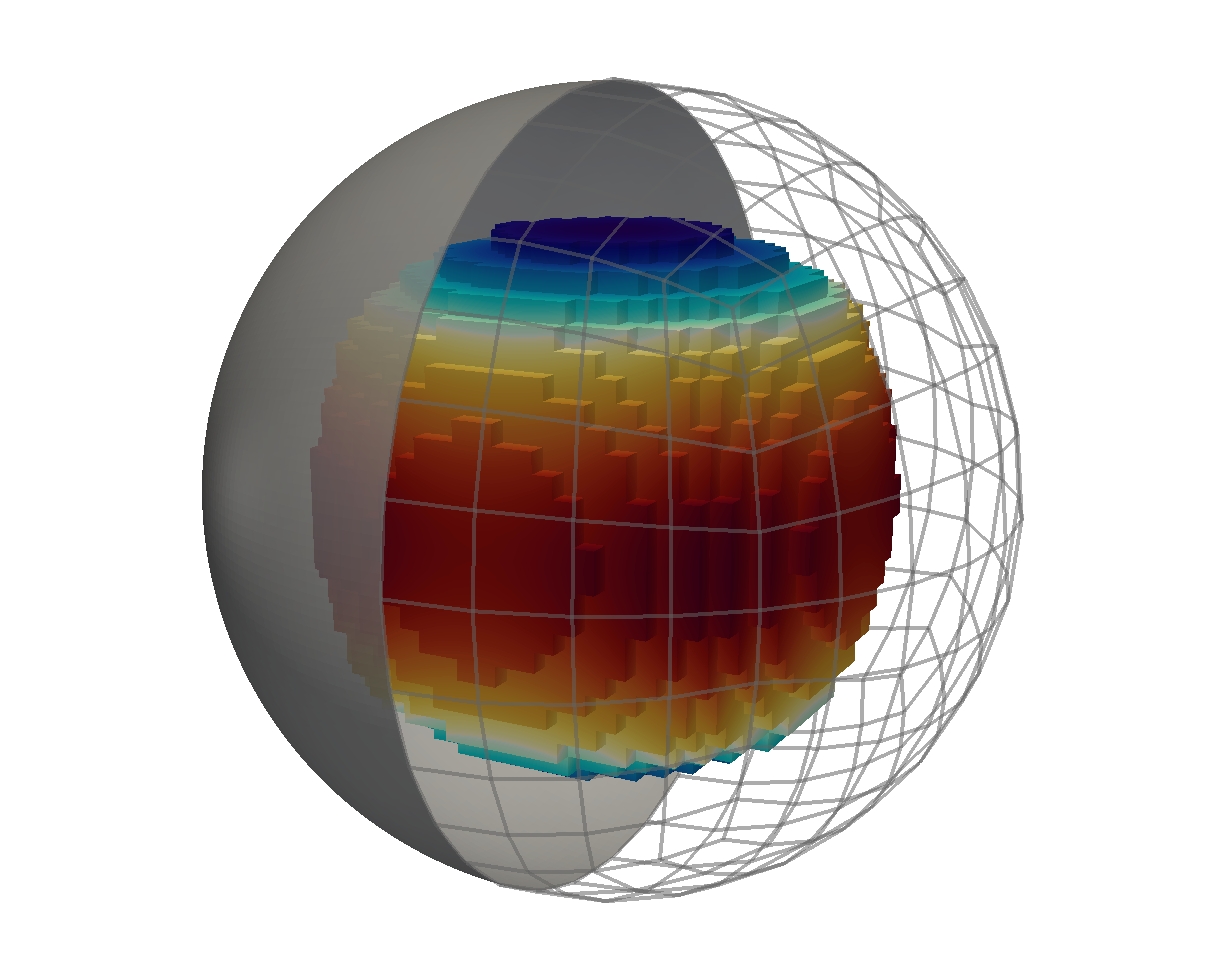}
	\vspace{-0.5cm}
  \caption{Boundary mesh and potential evaluation according to \eqref{eq::repform} in the interior (Laplace problem), see also \cite{Doelz_2017aa}.}\label{fig::potential}
\end{figure}

There exist different such representations, each with their specific strengths and drawbacks. As a simple example, the so called indirect representation formula for the Laplace Dirichlet problem 
\begin{align}
\begin{aligned}
    -\Delta u & = 0\qquad\text{in }\Omega \text{ or } \Omega^\text{C}\\
    u|_{\Gamma}
    & = g\qquad\text{on }\Gamma,\label{problem::helmholtz} 
\end{aligned}
\end{align}
is given by
\begin{align}
  u(\mathbf{x}) = \widetilde{\mathcal{V}}(\mu)(\mathbf{x})\coloneqq \int_\Gamma  \frac{1}{4\pi\vert\mathbf{x}-\mathbf{y}\vert} \mu(\mathbf{y}) ~\mathrm d \Gamma_{\mathbf{y}},\label{eq::repform}
\end{align}
on $\mathbb R^3\setminus \Gamma,$
where $\mu$ denotes some unknown density on the boundary. 
This density can be found by solving the arising variational problem: find $\mu$ in $X(\Gamma)$ such that
\begin{align}
  \left\langle \mathcal V(\mu),\nu \right\rangle_\Gamma = \left\langle g,\nu\right\rangle_\Gamma,\qquad \forall \nu \in X(\Gamma), \label{eq::varform}
  \intertext{ where $X(\Gamma)$ denotes an appropriate function space on the boundary and $\mathcal{V}$ is the restriction of $\widetilde{\mathcal{V}}$ to the boundary. Choosing a suitable 2D spline space~$\mathbb S(\Gamma),$ which can be defined patchwise as the 2D analogue of \eqref{eq:L2discreteconforming} mapped in the physical domain as in \eqref{eq:s1space}. This leads to the discrete formulation of finding a $\mu_h\in\mathbb S(\Gamma)$ such that}
  \left\langle \mathcal V(\mu_h),\nu_h \right\rangle_\Gamma = \left\langle g,\nu_h\right\rangle_\Gamma,\qquad \forall \nu_h \in \mathbb S(\Gamma). \label{eq::varformdisc}
\end{align}
After inserting the canonical basis of spline space $\mathbb{S}(\Gamma)$ we receive the linear system
\begin{align}
  \mathbf{Vw}=\mathbf g
\end{align}
where $\mathbf g$ encodes the discretization of the Dirichlet data given in \eqref{problem::helmholtz}, and $\mathbf w$ encodes the desired density function.
Plugging the discrete density back into \eqref{eq::repform}, one can evaluate the solution $u$ of \eqref{problem::helmholtz} in any given point $\mathbf{x}\notin \Gamma$, c.f. Fig.~\ref{fig::potential}.

This representation of the solution counteracts one of the greatest weaknesses of IGA FEM: where in many cases trivariate spline mappings need to be constructed by hand, a boundary element method can operate with a boundary representation (\emph{BRep}) only.
Moreover, it decreases the dimension of the problem: where FEM requires volumetric elements, BEM operates on the surface only, thus the elements are 2D elements in the sense of the reference domain.

It should be noted that the indirect representations are entirely insensitive to the choice of interior and exterior domain. Since the BEM operates on the boundary of the chosen domain only, this makes a boundary element approach an excellent choice to solve exterior problems, since neither bounding boxes nor DOF intensive exterior discretizations are required.

\begin{table}
  \centering
  \caption{Average magnetic field gradient and inhomogeneity factor before and after the optimization.}
  \label{tab:results}
  \centering
  \begin{tabular}{| l | c | c |}
    \hline 
    & $\tau_\mathrm{av}$ & $\epsilon$ \\ \hline
    {CST (3D, original geometry)}        &  $-237\,\mathrm{T/m}$ & 0.0503  \\
    {CST (3D, optimized geometry)}    &  $-266\,\mathrm{T/m}$ & 0.0201 \\ \hline
    {Improvement}    & $12.2\%$ & $60.0\%$ \\ \hline\hline
    
    {GeoPDEs (2D, original geometry)}  & $-240\,\mathrm{T/m}$ & 0.0477   \\
    {GeoPDEs (2D, optimized geometry)}      & $-282\,\mathrm{T/m}$ & 0.0122   \\ \hline
    {Improvement}    & $17.5\%$ & $74.4\%$ \\ \hline
  \end{tabular}
\end{table}

However, a BEM approach poses different challenges.
Due to the global nature of the integral term the matrices arising from \eqref{eq::varformdisc} are densely populated.
This turns out to be not as bad as it sounds: 
due to the decreased dimensionality fewer degrees of freedom are required, thus the linear systems become much smaller than in the case of FEM. Due to the approach via B-splines, this effect becomes even more notable. Even for real-world problems the arising linear systems are almost of sizes such that they can be solved in a dense representation by a desktop computer.
Moreover, efficient compression techniques to deal with the discrete linear systems are available. 
Such techniques are well understood and ready for industrial applications. An introduction to and comparison of compression methods can be found in \cite{Harbrecht_2013aa}.

An introduction into the realm of isogeometric BEM in the context of the Helmholtz equation which elaborates on everything mentioned above and explains the strength and drawbacks of the combination of IGA and BEM is given by \cite{Doelz_2017aa}; a first computational approach for the Maxwell case was also recently proposed~\cite{Simpson_2017aa}.
 
\section{Conclusions}
Isogeometric Analysis can be interpreted as a Finite Element Method that generalizes the set of basis functions from polynomials to B-splines and rational B-splines. This choice guarantees several advantages, from the exact parametrization of geometries defined via \gls{cad} to a higher accuracy per-degrees-of-freedom. It also allows for solution fields with higher smoothness. These reasons have made \gls{iga} a successful topic in recent years, particularly in the setting of mechanics and fluid simulation.

In this review article we show the benefits of \gls{iga} in the context of computational electromagnetics by presenting several 2- and 3-dimensional simulation examples. Its application in the context of shape optimization and uncertainty quantification is of particular interest given the possibility of deforming the domain and the mesh at the same time by means of a few control points. Finally, \gls{iga} shows great potential when combined with the boundary element approach since it eliminates the necessity of constructing trivariate parameterizations.

\section*{Acknowledgment}
This work is supported by the German BMBF in the context of the SIMUROM project (grant nr. 05M13RDA), by the DFG (grant nr. SCHO1562/3-1 and KU1553/4-1) and by the 'Excellence Initiative' of the German Federal and State Governments and the Graduate School of CE at TU Darmstadt.

\section*{Authors Name and Affiliation}
Zeger Bontinck, Jacopo Corno, Herbert De Gersem, Stefan Kurz, Andreas Pels, Sebastian Schöps, Felix Wolf from the Institut für Theorie Elektromagnetischer Felder and the Graduate School CE, Technische Universität Darmstadt, Darmstadt, Germany.\\[0.3em]
Carlo de Falco from MOX -- Modellistica e Calcolo Scientifico, Dipartimento di Matematica, Politecnico di Milano, Milano, Italy.\\[0.3em]
Jürgen Dölz from the Departement Mathematik und Informatik, Universität Basel, Basel, Switzerland.\\[0.3em]
Rafael Vázquez from the Institute of Mathematics, École Polytechnique Fédérale de Lausanne, Lausanne, Switzerland.\\[0.3em]
Ulrich Römer from the Department of Mechanical Engineering of Technische Universität Braunschweig, Braunschweig, Germany.\\[0.3em]
~\\
Corresponding author: Sebastian Schöps (\url{schoeps@gsc.tu-darmstadt.de})
\end{document}